\documentclass{article}
\usepackage{graphicx}
\usepackage{amsmath}
\usepackage{amsfonts}
\usepackage{amssymb}

\begin{document}

\title{Physicists in times of war}
\author{Bert Schroer\\CBPF, Rua Dr. Xavier Sigaud 150 \\22290-180 Rio de Janeiro, Brazil\\and Institut fuer Theoretische Physik der FU Berlin, Germany}
\date{December 2005}
\maketitle
\begin{abstract}
Though the majority of physicists would probably not support preemptive wars,
nuclear and other weapons of mass destruction would not exist without their contributions.

Einstein's anti-militaristic position has been well-documented and the present
essay recalls the role of some contemporary and past physicists on this issue.
The idea that the rationality of scientific thought is a reliable antidote
against supporting wars in order to achieve political or ideological aims was
neither correct in the past nor is it presently valid. In the physics
community there always existed a minority of supporters of wars of conquest of
territory, domination of people beyond the borders or regime change. The
``preemptive'' war for the US hegemony in the middle east has given the
problem of ``physicists in times of war'' new actuality, especially on the
third anniversary of its start with a country slipping into civil war and
culprit of this bloody mess talking about the fight for victory.

One of the most perplexing appologists of the agressive war of Nazi-Germany
against ``the Bolshevist peril'' has been the physicist Pascual Jordan whose
interesting scientific and controversial political biography is the main issue
of this essay.
\end{abstract}

\section{Introductory remarks}

As a result of a significant coincidence 2005 was not only the centennial year
of Einstein's greatest discoveries but it also marks the 60 year commemoration
of the end of the second world war, which was perhaps one of the greatest
man-made tragedy and certainly 20$^{th}$centuries darkest episode.

It is well-known that Albert Einstein, the most eminent physicist after Isaac
Newton, was also a renown anti-nationalist, anti-militarist and influential
champion of liberal humanitarian causes.

In this context the fate of the physicist Herbert Jehle comes to one's mind.
Herbert Jehle \cite{Je} was born 1907 in Stuttgart Germany as the son of
General Julius von Jehle. During his post-doc physics studies in the UK
1933-34, his pacifist views brought him in contact with the Quakers. Around
1936 he declined offers to work in the German rearmament and even turned down
an academic position in Berlin, thus following in the footsteps of his hero
Dietrich Bonhoeffer who renounced his university position 2 years before in
order to fight full-time the Nazification of the Lutheran church (Bonhoeffer
was hanged in Flossenburg just before the Americans arrived there). Jehle
refused military service 1940 before a Gestapo tribunal in France and was
interned in several Nazi-controlled concentration camps in Vichy France where
he barely escaped death by starvation before (with the outside help of a
fellow physicist and Quaker Sir Arthur Eddington) he got in 1941 to the US. He
participated in the postwar conferences where Feynman presented for the first
time his ideas and calculations to a wider public. Jehle pointed out some
interesting connections between the path integral and some of Dirac's ideas.
It is interesting to mention Jehle's life story in the present context because
it shows that a military family background of a person is not an insuperable
obstacle in the process of recognizing that war is one of the worst scourges
of mankind.

Several physicists among the many scientist refuges from Nazi-occupied Europe
left the Manhattan project and started their anti-nuclear weapon political
work as soon as Nazism was defeated. They realized that even a democratic
system could not provide any guaranty against the use of such weapons of mass
destruction but their influence was too weak to prevent their use by the
political and military elite in power against civilians\footnote{Although the
use of the nuclear bomb in Hiroshima and particular in Nagasaki fulfills the
definition of what constitutes a terroristic act, it is not clear that a mere
public demonstration of its destructive power could have led to a surrender of
Japan and in this way saved civilian lives. It is only in connection with the
recent agressive hegemonial policies and the arrogant unilateralism of the
Bush administration and its double standards concerning the nonproliferation
treaty (NPT) that these old episodes again enter the present discussions.}.
The nuclear disarmament activities of physicists as Leo Szilard and Joseph
Rotblat (the Nobel peace price winner of 1995) contributed to an improvement
of the public image of scientist and their moral responsibility.

The theme ``physicists in times of war'' has gained new topicality as a result
of the US doctrine of preemptive strikes adopted by the Bush administration.
This bellicose ideology has led to the Iraq war which according to all
criteria violates international law; in addition this war contradicts all
historical criteria for the bellum justum, including the prerequisites just
war in the sense of Saint Thomas of Aquin. During the last three years while
visiting many universities in Europe and Brazil I have never met any academic
person who, independent of his political affiliation did not consider this as
an illegal act. This in part explains my profound disappointment when some
weeks before the invasion of Iraq my attempts to encourage my US colleagues to
start a public anti-war campaign ended in failure. When Ed Witten informed me
that he is actually in favor of a war for regime change, I painfully realized
that I was still living in the past. We had a couple of email exchanges and
when I realized that I was not able to change his opinion that this was a good
idea, I reminded him that often the way into hell is paved by good intentions. 

Evidently times had changed since my 10 year stay in the US during the 60s. In
those days, as a result the cold war situation, there were no noticeable
differences in opinions on matters of peace and war in academic circles inside
and outside the US. If even somebody as Ed Witten, a champion of moderation
and of just and peaceful solutions of conflicts, is a supporter of the US
invasion, the deep split of parts of the US physics community with the rest of
the worlds physics community becomes palpable. There is no satisfaction in
pointing out that the present situation of death, destruction and the still
deteriorating chaos was foreseeable and in fact predicted; because both
blindness and hubris as well as cantankerousness are not human traits one can
be proud about.

The inability to see the obvious consequences of such a bellicose act reveals
a state of deep alienation with the rest of the world. There is no question
that the image of the US as a civilized member of the nations of this world
has been seriously damaged by the Iraq war and its present policies. The
support these policies still receives from some governments of Nato member
states cannot hide the fact that the majority of Europeans together with the
rest of the world are deeply disturbed by the US prepotency, hubris and
genuine ignorance about the world outside its borders. The creation of a
Guantanamo Gulag, the arrogance in the hegemonic redefinition of what
constitutes torture and kidnapping, the most degrading and gruesome behavior
as an occupation power has led to a worldwide image in which the US government
appears closer to represent some sort of Nazi-``Wiederg\"{a}nger'' than a
beacon of western civilization. If its aim is to loose more sympathies and
recruit more enemies, it could not have been more efficient.

The future scenario of a world-wide battle field is already in preparation. A
recent quotation from the Pentagon's four-yearly strategy review which sets
out a plan for prosecuting what the Pentagon describes in the preface as ``The
Long War'' (which replaces the ''war on terror'') reads:.

``\textit{Long duration, complex operations involving the US military, other
government agencies and international partners will be waged simultaneously in
multiple countries round the world, relying on a combination of direct
(visible) and indirect (clandestine) approaches,'' the report says. ''Above
all they will require persistent surveillance and vastly better intelligence
to locate enemy capabilities and personnel. They will also require global
mobility, rapid strike, sustained unconventional warfare, foreign internal
defence, counter-terrorism and counter-insurgency capabilities. Maintaining a
long-term, low-visibility presence in many areas of the world where US forces
do not traditionally operate will be required}.''

A poignant comment which fits such ambitions came from the late Peter Ustinov:

\textit{Terrorism is the war of the poor and war is the terrorism of the rich.}

The present situation cannot be explained in terms of a spontaneous reaction
to the 9.11. terrorism. To push the world to the brink of a ``war of
cultures'' is not possible without the politics based on a culture of war.

It seems that the planners of the ''The Long War'' are counting on a certain
apathy of the civil society in the US. Indeed, looking into US media one gets
the impression that even in those publications where the actions of the Bush
administration are criticized the present situation is more thought of in
terms of some neocon nuisance (something to be mentioned in the inner pages of
newspapers, i.e. nothing to really worry about) rather than a worldwide
destabilization, . It is deeply worrisome to see the foundations of the NPT
(the nuclear proliferation treaty) and the struggle for nuclear disarmament of
several generations of physicists (including very prominent US physicists) is
in the process of being wrecked by the present US government. There are
ominous reports about secret activities concerning the construction of a
completely new generations of nuclear weapons. The legal or moral grounds on
which the US will continue its campaign against Teheran if it rewards a
country like India which refused to sign the NPT with nuclear armament and
supports the secret development of its own construction of a new generation of
nuclear weapons is rapidly evaporating. The clear message to the rest of the
world is that of flagrant double standards; the hegemon defines his own rules
and coerces the others into obeying the existing ones (if deemed necessary
even with military threats). To understand the bad anti-US feelings this
creates in the world it is not necessary to create an extra government fact
finding position for to study and combat this phenomenon.

The idea that out of a sudden an unpleasant government with a disgusting point
of view comes to power may be convenient in avoiding to confront the
situation, but it is quite far from reality. Such changes rarely happen
suddenly, they need many smaller steps; a nation with a strong democratic
traditions as the US does not suddenly change into the role of an apologist or
revisionist of torture and kidnapping. Between its crucial role as one of the
liberators of Europe and the present inglorious situation of having destroyed
a country (which against better knowledge it insinuated to be implicated in
the twin tower massacre), killed an innumerable number of its citizens and
permitted the pillaging of their heritage, there are many episodes in which
its soul got lost, every time a bit more.

There are the many coups against democratically elected governments in the
Americas and the support of dictators as Samoza and Pinochet and that of death
squads in Nicaragua, San Salvador and Guatemala, the support of right-wing
paramilitary responsible for the assassination of Bishop Romero, the
``collateral damage'' of several thousand killed civilians in the capture of
Noriega in Panama City, the mining of Nicaraguan harbours, the financing of
torture schools where most of the Latin American torturers who served the US
supported military dictatorships learned their bloody trade, and last not
least the CIA financing of the Ben Laden led Al Qaida terrorist base during
the time when they directed their throat-cutting terror against the red army
in Afghanistan. As long as the US had to contain another superpower whose
methods in controlling its hemisphere of influence were even worse, these
discrepancies between the democratic ideal and the undemocratic practice were
irritating because they were caused by the governments of a country which sees
itself as a bulwark of freedom and democracy. With the vanishing of the Soviet
Union and the recent ascend of the neo-conservative right in the US the
situation changed completely. Double standards with respect to the world
outside the US and the substitution of the force of the law by the law of
force became the hall-mark of a neo-imperialist unilateralism, of which the
armed raid and occupation of the Iraq\ is the first bloody act with new
threats already in the wing. Whereas in previous times aggressive wars were
usually started by regimes which also had a dictatorial grip on power over
their own population, the new situation shows that under certain conditions
democracy inside a country can coexist with criminal actions outside its
borders. For this porpose the US government recruits its professional army
from its underpriviledged classes whose entrance into the army is often the
only way to obtain a slice of that liberty (from poverty) and democracy. How
such triggerhappy merceneries can convince the middle eastern people of the
blessings of democracy and liberty remains a secret of Bush and his neo-cons.
In order to minimize such inconsistencies the bush administration deemed it
important to redefine certain notions of what constitutes torture
($\rightarrow$rigorous interrogation) and kidnapping ($\rightarrow$rendition)
in case of suspicious non US citizens who it decides to consider as enemy
combatants outside any legal protection. It is a privilege of a hegemon to be
able to do this, he only has to make sure that his own population excepts the
new definitions, the compatibility with international law and human rights is
irrelevant. Just imagine the consequences if an US citizen would be kidnapped
by a foreign secret service and flown to another country!

The present situation begs the question why the tradition of western
civilization and democracy is not a reliable safeguard against policies which
lead to death and destruction. Limiting one's attention to the present
disastrous policies of the worlds superpower is a too narrow Ansatz, for a
profound insight it would be necessary to embark into a critical analysis of
modernity. The central problem and the key to many other human calamities has
been to understand what civilisatory defect of modernity has led to Auschwitz.
Attempts to relate this and other calamities of mankind to a special national
characters or intrinsic badness of certain ethnic groups have shown to be a
blind allay. The present attempts of Bush and the neocons to divide people
into the good guys (naturally we) and bad guys (the others) shows that
demagoguery does not stop at the doors of democracy. We are living in a world
where the oppressed of yesterday cannot rely on a special protection against
becoming the oppressors of today.

A more realistic attempt to understand this problem is to look at human nature
and realize that we are burdened with a potentially dangerous legacy which we
all share. Below a layer of civilization there are much older subconscious
regions of our brain which were important for our survival before we reached
the homo sapiens state, but which by irrational intrusions within the
enlightened rational parts of our civilized conscious may become activated and
unfold its destructive power. Most religions have created strong tabus and
commandments against these dangerous potentialities; the
Judeo-Christian-Muslim concept of humans as fallible and sin-prone creatures
is a recognition of the problem. But in situations of unexpected danger and
great social upheavals these taboos may loose their force and people succumb
to the hypnotic spell of demagogic leaders. The most difficult problem in
avoiding human catastrophes is to strike a delicate balance between upholding
strong taboos at times when a situation is running out of control (with the
obligation to call a spade a spade) and the comprehension afterwards that we
are dealing with a general problem of the human condition which a priori has
no special relation to a particular nation, race or culture. In relation to
the problem at hand, a fundamental critique of present neocon US policies does
not constitute anti-Americanism but the allegation that US-Americans have an
uncorrectable craving for world-dominating certainly would do just that.

Here the meticulous avoidance of double standards is essential. To give an
illustration how double standards have entered our everyday life one only has
to look at laws punishing anti-Semitism in some European countries. After the
holocaust rigorous laws have been introduced especially in Germany and Austria
(correct so), but to publish or re-print defamatory cartoons and statements
against the semitic majority of Arabs and Islam is protected under the freedom
of expression law (not so correct); the recent problem with anti-Arab and
anti-Muslim cartoons could have been avoided by extending the existing laws
against all kinds of defamations. Double standards actually do not prevent
anti-Semitism, in fact they are probably making an existent problem worse.

When we see the mass psychosis channeled through choreographed mass events in
e.g. old Nazi films, we find it hard to believe that our fellow human beings
in the past really acted in this way. Performances in the style of Mussolini
and Hitler appear to us outright ridiculous and chaplinesque and we seem to be
immune against that kind of racial ideology and propaganda for wars of
territorial extensions and domination of other races and cultures. The ascend
of the neoconservatives in the US with their fascist doctrine of preemptive
wars shows however that the belief that a democratic way of organizing society
alone can eliminate this problem is a delusion. The economic and military
hegemon only needs the support of big business, patriotic or submissive media
at the start of the war and an apathetic public; neocolonialism and high-tech
wars do not need fanatic masses. Democracy does not make the world less
dangerous but at least there is the chance of mistakes being corrected, as
long as the democratic will is not paralyzed by fear or stifled by emergency laws.

It is not possible to avoid wars and stop production of weapons of mass
destruction without creating strong rules and taboos. But they are without
moral strength if their enforcement leads to double standards. The times when
physicists were in the forefront of the struggle for a world without nuclear
arms are long gone.

Perhaps the most important contribution to a post Auschwitz fundamental
analysis of modernity comes from the writings of Theodor Adorno \cite{Wiki}.
The part which is relevant for arts and sciences is his ``Dialectics of
Enlightenment'' (with Max Horkheimer), his later essays on ``Kant's criticism
of pure reason'' as well as on ``Negative Dialectics''\footnote{These writings
had an enormous impact on the postwar philosophy and sociology in Germany but
(probably because they use all the resources of the German language and
philosophy) they have not played a comperable role in the anglo-saxon cultural
sphere.}. Faced with the unfolding events of the Holocaust in 1947, the work
``Dialectics of Enlightenment'' begins with the following remarkable passage:

\textit{Enlightenment, understood in the widest sense as the advance of
thought, has always aimed at liberating human beings from fear and installing
them as masters. Yet the wholly enlightened earth is radiant with triumphant calamity.}

Adorno's focal point is the mechanism by which rationality and enlightenment
can turn into irrationality. He illustrates his ideas mainly in arts and philosophy.

An essay about physicists and wars would suffer from a serious omission
without mentioning an eminent historical figure who has become the tragic
epitome of a scientists support of wars of aggression. I am thinking here of
Pascual Jordan who was together with Born and Heisenberg one of the
discoverers of quantum theory and the main protagonist of quantum field
theory. As a result of his bellicose and nationalistic stance he took right
after the first world war against the treaty of Versailles, he entangled
himself very much with the Nazi ideology. As a true believer in Heraklit's
dictum ``war is the father of all things'' he defended the idea that without
what he considered as the cleansing effect of war, mankind is condemned to
stagnation. This and his conviction that the ``Bolshevist peril'' had to be
eradicated drove him into the arms of the Nazis. This ideological support
remained one-sided since they never rewarded his backing by rewarded him with
a leading position in their weapons research program as they did in many other
cases. During the 30's, after the Nazis took power, he became increasingly
isolated even within the German physics community. His publications (mainly in
Zeitschrift fuer Physik\footnote{In the words of Peter Bergmann, a publication
in Z. f. Ph. after 1934 was tantamount to a first class burial.}) did not
receive the same attention as his earlier papers when he still had close
scientific contacts with Born, Heisenberg, Pauli, Wigner, Klein and von
Neumann. This explains why several important contributions by Jordan which
were ahead of times\footnote{E.g. the confirmation of magnetic monopole
quantization by a pure algebraic argument (using the formalism of exponential
line integrals which he introduced shortly before) and the observation of the
two-dimensional Bosonization/Fermionization.} went unnoticed. Illustrating the
topic ``physicists in times of war'' with a controversial figure from the past
may lack the tension caused by witnessing an ongoing human tragedy, but as a
result of the large distance in time, the context and the motives for a
particular belligerent behavior appear in a clearer light and are less subject
to future revisions.

Jordan's life is well-documented and serves as an interesting illustration
that even a brilliant independent scientific mind is not protected against
adopting an antihuman destructive political position. After the war he found
himself in a situation of a damaged reputation having lost the support of most
of his former friends and colleagues. This led to the unique and somewhat
anomalous situation that the co-discoverer of quantum mechanics and the
protagonist of quantum field theory was ignored by the Nobel prize committee;
instead Pascual Jordan became the tragic ``unsung hero of quantum field
theory'' \cite{Schweber}\cite{Darrigol}.

This essay continues with a short presentation of Jordan's interesting
biography with emphasis on those points which are relevant to the theme of
this essay.

\section{The case of Pascual Jordan, how the protagonist of quantum field
theory got himself entangled with the NS-regime}

There are not many physicists in whose biography the contradictions of human
existence, the proximity of glorious scientific achievements and disturbing
human weaknesses in the face of the great catastrophe of the 20th century, are
as starkly reflected as in the personality of Pascual Jordan\footnote{The
original title ``Pascual Jordan, Glory and...'' has been changed since
although the birth of quantum theory represents one of the most glorious
epochs in physics, Jordan himself remained ``the unsung hero'' among the
creators of that theory \cite{Schweber}.}.

Born on October 18, 1902 in Hannover of mixed German-Spanish ancestry, he
became (starting at age 22) a main architect of the conceptual and
mathematical foundations of quantum theory and the protagonist of quantum
field theory. Pascual Jordan owes his Spanish name to his great grandfather
Pascual Jorda, who came from the Alcoy branch (southern Spain) of the noble
Jorda family with a genealogy which can be traced back to the 9th century.
After the British-Spanish victory of Wellington over Napoleon, the family
patriarch Pascual Jorda settled in Hannover where he continued his service to
the British crown as a member of the \textquotedblleft
Koeniglich-Grossbritannisch-Hannoverschen Garde-Husaren
Regiments\textquotedblright\ until 1833. Every first-born son of the Jordan
(the n was added later) clan was called Pascual \cite{Ehlers}.

There is no doubt that Pascual Jordan took the lead in the formulation of the
conceptual and mathematical underpinnings of \textquotedblleft Matrix
Mechanics\textquotedblright\ in his important paper together with Max Born
\cite{B-J} submitted on 27. September 1925 (3 months after the submission of
Heisenberg's pivotal paper!) entitled \textquotedblleft Zur
Quantenmechanik\textquotedblright. His mathematical preparation, particularly
in the area of algebra, was superb. He had taken courses at the G\"{o}ttingen
mathematics department from Richard Courant and became his assistant (helping
in particular on the famous Courant-Hilbert book on mathematical methods in
physics); through Courant he got to know Hilbert before he met the 20 year
older Max Born, the director of the theoretical physics department of the
G\"{o}ttingen university. By that time Jordan already had gained his physics
credentials as a co-author of a book which he was writing at that time
together with James Franck \cite{Franck}.

After Max Born obtained Heisenberg's manuscript, he tried to make sense of the
new quantum objects introduced therein. While he had the right intuition about
their relation to matrices, he felt that it would be a good idea to look for a
younger collaborator with a strong mathematics background. After Pauli
rejected his proposal (he even expressed some reservations that Born's more
mathematically inclined program could stifle Heisenberg's powerful physical
intuition), Jordan volunteered to collaborate in this problem \cite{Pais}%
\cite{Jammer}. Within a matter of days he confirmed that Born's conjecture was
indeed consistent. The Born-Jordan results made Heisenberg's ideas more
concrete. Probably as a consequence of the acoustic similarity of pq with
Pascual, the younger members of the physics department (the protagonists of
the ``Knabenphysik'') in their discussions often called it the Jordan
relation. Max Born became Jordan's mentor in physics. Jordan always maintained
the greatest respect for Born which withstood all later political and
ideological differences.

The year 1925 was a bright start for the 22-year-old Jordan. After the
submission of the joint work with Max Born on matrix mechanics, in which the
p-q commutation relation appeared for the first time, there came the famous
''Dreimaennerarbeit''\cite{B-H-J} with Born and Heisenberg in November of the
same year\footnote{Appearantly while writing up the last section on the
oscillator description of radiation, Jordan already had set his thoughts on
the more daring generalization of quantizing nonrelativistic matter $\psi$
waves for the only purpose to quantize them. It apparently did not occur to
him that what he thought to be a classical theory was in fact Schr\"{o}%
dinger's formulation of QM \cite{Darrigol}. For some time his colleagues did
not accept that something which was already a quantum theory should be
subjected to a second quantization.}, only to conclude the year's harvest with
a paper by him alone on the ``Pauli statistics''. Jordan's manuscript
contained what is nowadays known as the Fermi-Dirac statistics; however it
encountered an extremely unfortunate fate after its submission because it
landed on the bottom of one of Max Born's suitcases (in his role as one of the
editors of the Zeitschrift fuer Physik) on the eve of an extended lecture tour
to the US, where it remained for about half a year. When Born discovered this
mishap, the papers of Dirac and Fermi were already in the process of being
published. In the words of Max Born \cite{Born}\cite{Schucking} a quarter of a
century later: ''I hate Jordan's politics, but I can never undo what I did to
him......When I returned to Germany half a year later I found the paper on the
bottom of my suitcase. It contained what one calls nowadays the Fermi-Dirac
statistics. In the meantime it was independently discovered by Enrico Fermi
and Paul Dirac. But Jordan was the first''\footnote{In a correspondence with
Stanley Deser, Stanley added a light Near East touch by remarking that without
Max Born's faux pas the Fermions would have been called ``Jordanons''.}. In
Jordan's subsequent papers, including those with other authors such as Eugene
Wigner and Oscar Klein, it was always referred to as ``Pauli
statistics''\ because for Jordan it resulted from a straightforward
algebraization of Pauli's exlusion principle.

From later writings of Born and Heisenberg we also know that Jordan
contributed the sections on the statistical mechanics (or rather kinetic gas
theory) consequences to the joint papers on matrix mechanics. This is not
surprising since the main point in his 1924 PhD thesis was the treatment of
photons according to Planck's distribution whereas thermal aspects of matter
were described according to Boltzmann. He continued this line of research by
introducing the ``Stosszahlansatz''\ for photons and using for electrons and
atoms the Bose statistics \cite{thermal}\footnote{This paper was submitted
simultaneously with another paper in which Jordan coined the term
``Pauli-Principle''; but the relation to statistics was only seen later.}
which brought praise as well as criticism by Einstein's and led to an
unfortunately largely lost correspondence. In the following we will continue
to mention his scientific contributions in the biographical context and
reserve a more detailed account about their scientific content to the next section.

The years 1926/27 were perhaps the most important years in Jordan's career in
which he succeeded to impress his peers with works of astonishing originality.
The key words are Transformation Theory \cite{Trans}\cite{Trans2} and
Canonical Anti-Commutation Relations \cite{Gas}. With these discoveries he
established himself as the friendly competitor of Dirac on the continental
side of the channel and in its printed form one finds an acknowledgment of
Dirac`s manuscript\footnote{In those days papers were presented in a factual
and very courteous style; however verbal discussions and correspondences were
sometimes more direct and less amiable (e.g. see some published letters of
Pauli \cite{Pais}\cite{Schweber}).}. As an interesting sideline, one notes
that in a footnote at the beginning of Jordan's paper about transformation
theory Jordan mentions a ``very clear and transparent treatment''\ of the same
problem in a manuscript by Fritz London, a paper which he received after
completing his own work and which was published in \cite{London}. This
statement of Jordan's was not just the standard modesty of those days, but
really the truth. A glance into this largely overlooked paper confirms
Jordan's high praise; London's version of transformation theory was by far the
most cleare and advanced of the three presentations of transformation
theory\footnote{One usually links London's name with his work on the hydrogen
molecule and his studies of superconductivity theory and overlooks his
brilliant contributions to mathematical physics. He was the first physicist
who introduced Hilbert spaces into quantum mechanics and he had the clearest
vision about the involved operators (which he called ``rotations in Hilbert
space``) and the equivalence between Heisenberg's and Schr\"{o}dinger's
formulations \cite{London}.}. Most physicists are more familiar with Dirac's
notation (as the result of his very influential textbook whose first edition
appeared in 1930). Jordan's most seminal contribution is perhaps his 1927
discovery of ``Quantization of Wave Fields''\ which marks the birth of QFT.

Pascual Jordan was brought up in a traditional religious surrounding. At the
age of 12 he apparently went through a soul-searching fundamentalist period
(not uncommon for a bright youngster who tries to come to terms with rigid
traditions) in which he wanted to uphold a literal interpretation of the bible
against the materialistic Darwinism (which he experienced as a ''qu\"{a}lendes
Aergernis'', a painful calamity), but his more progressive teacher of religion
convinced him that there is basically no contradiction between religion and
the sciences. This then became a theme which accompanied him throughout his
life; he wrote many articles and presented innumerable talks on the subject of
religion and science.

At the times of the great discoveries in quantum field theory many of his
colleagues thought that the treaty of Versailles was unjust and endangered the
young Weimar Republic, but Jordan's political inclination went far beyond and
became increasingly nationalistic and right-wing. These were of course not
very good prerequisites for resisting the temptations of the NS movement, in
particular since the conservative wing of the protestant church (to which he
adhered\footnote{The oldest son of the family patriarch Pascual Jorda was
brought up in the Lutheran faith of his foster mother, whereas all the other
children born within that marriage were raised in the Catholic faith.})
started to support Hitler in the 30's; in fact the behavior of both of the
traditional churches during the NS regime belongs to their darkest chapters.
Hitler presented his war of aggression as a divine mission and considered
himself as an instrument of God's predestination (g\"{o}ttliche Vorsehung),
while almost all Christian churches were silent or even supportive.

Already in the late 20s Jordan published articles \cite{Schucking} (under a
pseudonym) of an aggressive and bellicose stance in journals dedicated to the
spirit of German Heritage; a characteristic ideology of right-wing people up
to this day if one looks at the present-day heritage foundations and their
political power in the US. It is unclear to what degree his more cosmopolitan
academic peers in G\"{o}ttingen knew about these activities. He considered the
October revolution and the founding of the Soviet Union as extremely worrisome
developments. One reason why Jordan succumbed to the NS-lure was perhaps the
idea that he could influence the new regime; his most bizarre project in this
direction was to convince the party leaders that modern physics, as
represented by Einstein and especially the new Copenhagen brand of quantum
theory, was the best antidote against the ``materialism of the Bolsheviks''.
This explains perhaps why he joined NS organisations at an early date when
there was yet no pressure to do so \cite{Wise}. In fact he apparently thought
that he could establish a link between ``the new order'' of the NS state and
the strange counterintuitive dynamics of the Copenhagen interpretation and its
new conquest of reality \cite{Cornwell}. Among all attempts to carry the power
of the new quantum physics beyond its range of validity, this was certainly
the most bizarre.

He of course failed in his attempts; despite verbal support\footnote{In
contrast to Heisenberg he did not directly work on any armament project but
rather did most of his military service as a meteorologist.} he gave to their
nationalistic and bellicose propaganda and even despite their very strong
anti-communist and anti-Soviet stance with which he fully agreed, the
anti-semitism of the Nazis did not permit such a viewpoint since they
considered Einstein's relativity and the modern quantum theory with its
Copenhagen interpretation as incompatible with their anti-semitic propaganda;
one can also safely assume that the intense collaboration with his Jewish
colleagues made him appear less than trustworthy in the eyes of the regime.

Jordan's career during the NS time ended practically in scientific isolation
at the small university of Rostock (his promotion to fill von Laue's position
in Berlin in 1944 was too late for a new start); he never received benefits
for his pro-NS convictions and the sympathy remained one-sided. Unlike the
mathematician Teichmueller, whose rabid anti-semitism led to the emptying of
the G\"{o}ttingen mathematics department, Jordan inflicted the damage mainly
on himself. The Nazis welcomed his verbal support, but he always remained a
somewhat suspicious character to them. As a result he was not called upon to
participate in war-related projects (as e.g. the uranium project) and spent
most of those years in scientific isolation. This is somewhat surprising in
view of the fact that Jordan, like nobody else, tried to convince the NS
regime that fundamental research should receive more support because of its
potential weapons-related applications; in these attempts he came closer to a
``star wars''\ propagandist of the Nazis than Heisenberg who headed the German
uranium program but never joined the party.

Jordan's party membership and his radical verbal support in several articles
got him into trouble after the war. For two years he was without any work and
even after his re-installment as a university professor he had to wait until
1953 for the reinstatement of his full rights (e.g. to advise PhD candidates).
When his friend and colleague Wolfgang Pauli asked him after the war:
``Jordan, how could you write such things?'' Jordan retorted: ``Pauli, how
could you read such a thing?'' Without Heisenberg's and Pauli's help he would
not have been able to pass through the process of de-nazification (in the
jargon of those days Jordan got a ``Persilschein'', i.e. a whitewash paper)
and afterwards to be re-installed as a university professor. In Pauli's
acerbic way of dealing with such problems: ``Jordan is in the possession of a
pocket spectrometer by which he is able to distinguish intense brown from a
deep red''. ``Jordan served every regime trustfully'' is another of Pauli's
comments. Pauli recommended Jordan for a position at the University of Hamburg
and he also suggested that he should keep away from politics and rather worry
about his pension.

Jordan did not heed Pauli's advice for long; during the time of Konrad
Adenauer and the big debates about the re-armament of West Germany he became a
CDU member of parliament. His speech problem (he sometimes fell into a
stuttering mode which was quite painful for people who were not accustomed to
him) prevented him from becoming a scientific figurehead of the CDU party. At
that time of the re-armament issue there was a manifesto by the ``G\"{o}%
ttingen 18'' which was signed by all the famous names of the early days of the
university of G\"{o}ttingen quantum theory, including Max Born. Jordan
immediately wrote a counter article with the CDU party's blessing, in which he
severely criticized the 18 and claimed that by their action they endangered
world peace and stability. Max Born felt irritated by Jordan's article, but he
did not react in public against Jordan's opinion. What annoyed him especially
were Jordan's attempts to disclaim full responsibility for his article by
arguing that some of the misunderstandings resulted from the fact that it was
written in a hurry. But Born's wife Hedwig exposed her anger in a long letter
to Jordan in which she blamed him for ``deep misunderstanding of fundamental
issues''. She quoted excerpts from Jordan's books and wrote: ''Reines
Entsetzen packt mich, wenn ich in Ihren B\"{u}chern lese, wie da menschliches
Leid abgetan wird'' (pure horror overcomes me when I read in your books how
human suffering is taken lightly). Immediately after this episode she
collected all of Jordan's political writings and published them under the
title: ``Pascual Jordan, Propagandist im Sold der CDU'' (Pascual Jordan,
propagandist in the pay of the CDU) in the Deutsche Volkszeitung.

In the middle of the twenties the authors of the ``Dreimaennerarbeit'' were
proposed twice for the Nobel prize by Einstein, but understandably the support
for Jordan dwindled after the war. Nevertheless, in 1979 it was his former
colleague and meanwhile Nobel prize laureate Eugene Wigner who proposed him.
But at that time the Nobel committee was already considering second generation
candidates associated with the second phase of QFT which started after the war
with perturbative renormalization theory; there was hardly any topic left of
the first pioneering phase which was not already taken into account in
previous awards. Jordan did however receive several other honors, including
the Max-Planck-medal of the German Physical Society.

Although Jordan took (along with the majority of German physicists) a strong
position against those supporting the racist ``German Physics''\footnote{It
was Jordan's opinion that nationalistic and racist views had no place in
science; in his own bellicose style of ridicule (in this case especially
directed against nationalistic and racist stance of the mathematician
Bieberbach): ``The differences among German and French mathematics are not any
more essential than the differences between German and French machine guns''.}
and in this way contributed to their downfall, he defended bellicose and
nationalistic positions and he certainly supported Hitler's war of aggression
against the ``Bolshevik peril''. The fact that he was a traditional religious
person and that several of the leading bishops in the protestant church were
pro Hitler had evidently a stronger effect on him than his friendship with his
Jewish colleagues, who by that time had mostly left Germany (in some cases he
tried to maintain a link through correspondence).

In contrast to Pauli who contributed to the second post war phase of QFT and
always followed the flow of ideas in QFT up to his early death, Jordan's
active participation in QFT stopped around the middle of the 30s and it seems
that he did not follow the continuing development in that area. He turned his
attention to more mathematical and conceptual problems as well as to biology
\cite{Beyler} and psychology. His enduring interest in psychology was
presumably related to the psychological origins of his stuttering handicap
which prevented him from using his elegant writing style in discussions with
his colleagues and communications with a wider audience (one should keep in
mind that people at that time had lesser tolerance with physical and psychic
handicaps); this perhaps explains in part why even in the physics community of
the 30s his contributions are not as well known as they deserve to be. In fact
this handicap even threatened his Habilitation (which was a necessary step for
an academic career) in G\"{o}ttingen. Jordan was informed by Franck (with whom
he had coauthored a book) that Niels Bohr\footnote{One also should keep in
mind that the interest in psychology became a \textquotedblleft
fashion\textquotedblright\ among the Copenhagen physicists (notably Bohr and
Pauli).} had arranged a small amount of money for Jordan which was to be used
for getting some cure of his speech problem. Wilhelm Lenz (whose assistant
Jordan was for a short time after Pauli left) suggested to go to the famous
psychologist Adler. Jordan went to Vienna, but we only know that he attended a
lecture of Schr\"{o}dinger and criticized his wave mechanics from the
G\"{o}ttingen point of view; there is no record of meetings with Adler.

His increasing withdrawal from the mainstream of quantum field theory and
particle physics in the 30s may have partially been the result of his
frustration that his influence on the NS regime was not what he had expected.
After the defeat of Germany in 1945 his attempts to account for his membership
in the Nazi party as well as the difficult task to make a living with the
weight of his past NS sympathies (which cost him his position as a university
professor for the first two years after the war) seriously impeded his
scientific activities, although there is no indication that he was antisemitic
(his naive attempts to influence the NS regime to accept the work of Einstein
and the Copenhagen view of QM shows however that he was not able or did not
want to see the true nature of that regime).

Unlike the majority of the German population, for which the early Allied
re-education effort (which was abandoned after a few years) to rid society of
aggressive militaristic and racist ideas was a huge success so that the
subsequent change of US policy in favor of re-armament of West Germany ran
into serious opposition during the Adenauer period, Jordan did not completely
abandon his militaristic and rightwing outlook. In the 50s he joined the CDU,
a party which was closer to his opinions, thus ignoring Pauli's admonitions in
favor of political abstinence. Bellicose ideas and apologetic positions with
respect to wars of aggressions were not totally uncommon between Jordan's
contemporaries. For the post second world war generations statements like ''a
war is the normal way to accomplish something new in history'' (Jordan)
hopefully did not loose any of its outrageous content as a result that point
of views like this recently led some democratically elected governments into
what they euphemistically called a preemptive war. However it is also true
that among all physicists who put their knowledge at the disposition of
martial applications, Jordan was probably the most inefficient contributor.

All the protagonists of those pioneering days of quantum physics have been
commemorated in centennials except Pascual Jordan who, as the result of the
history we have described, apparently remained a somewhat ``sticky'' problem
despite Pauli's intercession by stating ``it would be incorrect for West
Germany to ignore a person like P. Jordan''. His postwar scientific activities
consisted mainly in creating and arranging material support (by grants from
Academies and Industry as well as from the US Air-force) for a very successful
group of highly motivated and talented young researchers in the area of
General Relativity who became internationally known (Engelbert Schuecking,
Juergen Ehlers,..) and attracted famous visitors especially from Peter
Bergmann's group (Rainer Sachs,....). .

Jordan died in 1980 (while working on his pet theory of gravitation with a
time-dependent gravitational coupling); his post war work never reached the
level of the papers from those glorious years 1925-1930 or his subsequent
rather deep pre-war mathematical physics contributions. In the words of Silvan
Schweber in his history of quantum electrodynamics, Jordan became the ``unsung
hero'' of a glorious epoch of physics which led to the demise of one of its
main architects.

It is however fair to note that with the exception of Max Born, Jordan's other
collaborators, especially von Neumann and Wigner, shared the bellicose kind of
anti-communism (in their case it probably had its roots in their experience
with the radical post World War I Bela Kuhn regime in the Hungarian part of
the decaying Habsburg empire). Wigner later became an ardent defender of the
Vietnam war. Although Jordan's early right wing anti-communist views
apparently posed no friction during the time of his collaboration with Wigner
and von Neumann up to 1936; his membership in the Nazi party finally damaged
that relation. Nevertheless Wigner was one of the few of Jordan's
contemporaries who, after receiving the Nobel prize in 1963 (together with
Maria Goeppert-Mayer and Hans Daniel Jensen) supported the idea to honor
Jordan's epoch-changing contributions to quantum physics notwithstanding his
political and human errors.

\textit{Note added:} After I placed this essay onto the server, two of my
colleagues from the US pointed out to me that my impression about lethargy and
complacency of the academic community in the US needed some corrections in
view of the fact that Prof. Walter Kohn's initiative. I had the bad luck to
encourage Ed Witten in an email to start an anti-war campaign and I
interpreted his negative answer in a too pessimistic way and resigned. I have
been an associate professor at the University of Pittsburg during the 60s
(with very good memories from those times) before I took a full professorship
from the FU-Berlin. After my retirement I went to Brazil. I sometimes look
into the internet sites of international newspapers, but I am somewhat
detached from the present mood in US academic institutions.

\end{document}